\documentclass[12pt]{article}
\usepackage{graphicx}
\usepackage{amsmath}
\usepackage{amssymb}
\usepackage{latexsym}

\newcommand{\br}{\mathbb R}
\newcommand{\bz}{\mathbb Z}

\newcommand{\cA}{{\cal A}}

\newcommand{\cH}{{\cal H}}
\newcommand{\cK}{{\cal K}}

\newcommand{\cT}{{\cal T}}
\newcommand{\cO}{{\cal O}}
\newcommand{\cM}{{\cal M}}
\newcommand{\cN}{{\cal N}}
\newcommand{\cD}{{\cal D}}

\newcommand{\kett}[1]{{\left|#1\right\rangle}}

\newcommand{\ket}[1]{{|#1\rangle}{}}
\newcommand{\bra}[1]{{\langle#1|}}

\newcommand{\tq}{\tilde{q}}

\newcommand{\tr}{\mbox{Tr}}

\newcommand{\nn}{\nonumber\\}

\makeatletter
\@addtoreset{equation}{section}
\def\theequation{\thesection.\arabic{equation}}
\makeatother

\begin{document}

\vskip 7mm
%%% Title page %%%%%
\begin{titlepage}
 
 \renewcommand{\thefootnote}{\fnsymbol{footnote}}
 \font\csc=cmcsc10 scaled\magstep1
 {\baselineskip=14pt
 \rightline{
 \vbox{\hbox{hep-th/0201175}
       \hbox{UT-986, UT-02-01}
       }}}

 \vfill
 \baselineskip=20pt
 \begin{center}
 \centerline{\LARGE \bf Orientifolds of $SU(2)/U(1)$ WZW Models} 

 \vskip 2.0 truecm

\noindent{ \large Yasuaki Hikida}\footnote{
E-mail: hikida@hep-th.phys.s.u-tokyo.ac.jp}
\bigskip

 \vskip .6 truecm
 {\baselineskip=15pt
 {\it Department of Physics,  Faculty of Science, University of Tokyo \\
  Hongo 7-3-1, Bunkyo-ku, Tokyo 113-0033, Japan} 
 }
 \vskip .4 truecm

 \end{center}

 \vfill
 \vskip 0.5 truecm

\begin{abstract}
\baselineskip 18pt

The orientifolds of $SU(2)/U(1)$ gauged WZW models are investigated.
In particular, we construct the new type orientifolds and identify their
geometries. 
We closely follow the analysis of D-branes in the $SU(2)/U(1)$ WZW
models, which was given by Maldacena, Moore and Seiberg.

\end{abstract}
 \vfill
 \vskip 0.5 truecm

\setcounter{footnote}{0}
\renewcommand{\thefootnote}{\arabic{footnote}}
\end{titlepage}

\newpage
 \baselineskip  18pt

%%%%%%%%%%%%%%%%%%%%%%%%%%%%%%%%%%%%%%%%%%%%%%%%%%%%%%%%%%%%%%%%%%%%

\section{Introduction}
\indent

In recent years the D-branes attract much attention to investigate
several aspects of superstring theories.
Much of the studies of D-branes have been done in the flat background
and it is natural to extend the analysis to the curved background.
In the case of WZW models, 
the studies of D-branes were started about ten years ago
\cite{ishibashi,cardy}. 
Recently the geometrical aspects of D-branes in WZW models were
investigated in \cite{KO,AS1,Gawedzki1,zuber,FFFS,stanciu} 
and the stability of D-branes was discussed in
\cite{alek1,bds,paw,alek2}.

The D-branes are described by so called boundary states in the conformal
field theory and we can analyze perturbatively. It is known that the
orientifolds can be also expressed by the crosscap states
\cite{CLNY,Pol} in the conformal field theory, however the orientifolds
have not been much investigated like D-branes.
Nevertheless it is important to study the orientifolds because
they can be used to study the string duality and
to construct the configuration with less supersymmetry.
The orientifolds of WZW models were studied in  
\cite{sagnotti1,sagnotti2,sagnotti3,sagnotti4,simple1,simple2} by the
algebraic methods.
Very recently, the geometrical aspects of the orientifolds of WZW models
were studied in \cite{oplane1,oplane2,oplane3,oplane4} and it is
interesting to apply to other configurations.     

In this paper we will consider the orientifolds in the parafermionic
theories, which are obtained by the $SU(2)$ WZW models gauged by the
$U(1)$ sector, e.g., $SU(2)/U(1)$ coset WZW models \cite{GQ}.
These models appear in superstring theories on some backgrounds.
For example, the near horizon geometry of NS5-branes can be
described by \cite{chs1,chs2,chs3}
\begin{equation}
 \br^{1,5} \times \br_{\phi} \times S^3~, 
\end{equation}
where $\br^{1,5}$ is the space parallel to the NS5-branes.
The radial direction $\br_{\phi}$ is given by the Liouville theory 
and $S^3$ can be described by the $SU(2)$ WZW models.
Moreover the parafermionic theories are used to describe the string
theory on two dimensional black hole \cite{2bh} or singular Calabi-Yau
manifolds \cite{GV,OV}, which can be described by the $SL(2,\br)/U(1)$
WZW models. 

The study of D-branes in $SU(2)_k/U(1)_k$ WZW models was started by
\cite{RS} in the context of Gepner models and the geometrical picture of
D-branes in the parafermionic theories was given in \cite{MMS}.
Some branes which preserve maximal symmetry  
can be obtained in the rational conformal
field theory by Cardy construction \cite{cardy}, which will be called as
A-type branes followed by \cite{MMS}.
The authors of \cite{MMS} also constructed new types of (B-type) 
boundary states which can not be obtained by Cardy construction.
It can be shown that the $\bz_k$ orbifold of parafermionic theory   
is T-dual to the original theory.
Therefore we can construct new type boundary states by applying $\bz_k$
orbifold and T-duality.
For example, in $U(1)$ WZW model, A-type boundary states represent the
branes satisfying Dirichlet boundary condition and B-type boundary
states represent the branes satisfying Neumann boundary condition.
In the parafermionic theory, the target space is given by the disk with
radius one. In the paper \cite{MMS}, A-type branes were determined as
D0-branes at the boundary of the disk and D1-branes connecting the
points of the boundary. Moreover B-type branes were determined
as D0-brane and D2-branes at the center of the disk. 
The recent developments about D-branes in coset WZW models were given in
\cite{Gawedzki2,Parvizi,Elitzur,coset,Ishikawa,Kubota,Forste,Nozaki}. 

In this paper, we investigate the orientifolds in the parafermion
theory. The Cardy type orientifolds are known 
\cite{sagnotti1,sagnotti2,sagnotti3,sagnotti4} and they will be called
as the A-type orientifolds. We construct the new type (B-type)
orientifolds by using $\bz_k$ orbifold and T-duality 
and identify the geometry of these
orientifolds. A-orientifolds are the $\cO$1-planes connecting the
opposite boundary
points and B-orientifolds are the $\cO$0-plane at the center of the disk
and $\cO$2-plane wrapping the whole disk.

The organization of this paper is as follows; in section 2, we review
the known results of orientifolds which are constructed by the crosscap
state technique. 
The spectrum can be read from the amplitudes between the boundary states
or the crosscap states and  
it is very important consistency condition that there
appears only integer degeneracies.
In section 3, we investigate the orientifolds of the simplest
case, namely, $U(1)$ WZW
models. We can construct B-type orientifolds from A-type ones even in
this case. The orientifolds of $U(1)$ WZW models are the well-known ones
in the one dimensional free bosonic theory; A-type ones can be
identified as $\cO$0-planes and B-type ones as $\cO$1-plane.
In section 4, we construct the orientifolds in the parafermionic theory.
As we said above, there are A-type and B-type orientifolds and we
examine the spectrum in the direct channel of the amplitudes.
The geometry of orientifolds can be seen by the spectrum or by
scattering with the closed string states as we explain in appendix B.
In section 5, we extend the analysis to the supersymmetric case and
obtain the results similar to the bosonic theory.
Section 6 is devoted to the conclusions and discussions.
In appendix A, the several functions and their modular
transformations are denoted.
In appendix B, we review the method to determine the geometry of branes
by scattering the closed string states \cite{MMS} and identify the
shape of orientifolds by applying this method.
\\

\noindent{\bf Note added:} 
In the revised version, the geometry of A-type orientifolds is
corrected and the spectral flow identification is reconsidered. 
The results agree with the ones in the recent preprint \cite{BH}.

%%%%%%%%%%%%%%%%%%%%%%%%%%%%%%%%%%%%%%%%%%%%%%%%%%%%%%%%%%%%%%%%%%%%%

\section{General Analysis of Crosscap States}
\indent

In the presence of orientifolds, the partition functions can be
obtained by the Klein bottle worldsheet in addition to the torus one. 
If we add the open string sector, we have to consider the M\"{o}bius
amplitudes and of course the annulus amplitudes. 
These amplitudes have to satisfy the constraints corresponding to the
modular invariant.
In this section, we use the rational conformal field theory and we
later analyze the specific models. 

The modular invariant condition of the torus amplitudes is well-known.
The torus amplitudes can be written as
\begin{equation}
 \cT (q)= \sum_{ij} \chi_i(q) Z_{ij} \overline{\chi_j (q)} ~,
\end{equation}
where we use $\chi_i (q)$ as the character of the representation $i$ and 
the moduli $q = \exp (2 \pi i\tau)$ and 
$\tq = \exp (2 \pi i (- 1 / \tau ))$.
The torus amplitudes must be invariant under the modular transformations,
in other words, $S$ and $T$ transformations
\begin{equation}
 \sum_{ij} S_{i'i} Z_{ij} S^{\dagger}_{~jj'}  = Z_{{i'} {j'}}~,~~
 \sum_{ij} T_{i'i} Z_{ij} T^{\dagger}_{~jj'}  = Z_{{i'} {j'}} ~.
\end{equation}
The famous ones are given by the charge conjugation modular invariant
$Z_{ij}=\delta_{i,j^*}$ and the diagonal modular invariant 
$Z_{ij} = \delta_{i,j}$.
The analysis of orientifolds is usually given in the charge
conjugation modular invariant, however we will use the diagonal modular
invariant in the subsequent sections.

When applying the orientifold projection, we have to add the Klein bottle
amplitudes. The orientifold operation is given by the combination of the
worldsheet orientation reversal ($\Omega:\sigma \to 2 \pi - \sigma$)
and the space $\bz_2$ isometry ($h$).  
Then the Klein bottle amplitudes are given by
\begin{equation}
 \cK (q) = \tr (\Omega h 
 q ^ {H_c}) = \sum_i \cK^i \chi_i (q) ~,
\label{KB1}
\end{equation} 
where we define $H_c = \frac{1}{2}(L_0 + \tilde{L}_0 - \frac{c}{12})$
 and $c$ is the central charge of the model.
The coefficients $\cK^i$ are the integers 
and they are related to 
how the fields behave under the orientifold operation, which must be
consistent with OPEs of the fields \cite{sagnotti4}. 
The modulus of the Klein bottle is $\tilde{\tau} = 2 i t$ and $S$
 transformation interchanges the direct channel and the transverse channel.
In the transverse channel, the Klein bottle amplitudes are given by the
overlaps between the states called as crosscap states 
\begin{equation}
 \cK (\tq) = ~_C\bra{C} \tq^{H_c} \ket{C}_C 
      =\sum_i (\Gamma^i)^2 \chi_i (\tq)~.
\label{KB2}
\end{equation}  
The coefficients $\Gamma^i$ are obtained by the one point
amplitudes with the closed string $i$ 
and the coefficients $\Gamma^i$ determine the crosscap states. 
The equations (\ref{KB1}) and (\ref{KB2}) are transformed by $S$
transformation each other. Therefore we obtain the constraint of the
coefficients as 
\begin{equation}
 \cK^i  = \sum_j(\Gamma^j)^2 S_{j}^{~i}~,
\end{equation}
which is the important condition to construct the crosscap states.

The open string sector is included in the presence of D-branes.
The annulus amplitudes are given by
\begin{equation}
 \cA_{ab}(q) = \tr _{\cH_{ab}} {q}^{L_0 - \frac{c}{24}} 
               = \sum_i A^i_{ab} \chi_i (q)  ~,
\label{A1}
\end{equation}
where $a,b$ are the labels of the boundary conditions, in other words, the 
labels of D-branes.
The coefficients $A^i_{ab}$ are the non-negative integers. 
The transverse channel can be obtained by the overlaps between
the boundary states \cite{ishibashi,cardy},
which describe D-branes, as
\begin{equation}
 \cA_{ab} (\tq) = ~_C \bra{B,a} \tq^{H_c} \ket{B,b}_C 
                = \sum_i B^i_a B^i_b \chi _i (\tq) ~,
\label{A2}
\end{equation}
where $B^i_a$ are the one point disc amplitudes with the boundary
condition $a$.
The $S$ transformation connects the equations (\ref{A1}) and
(\ref{A2}) and this constraint is known as the Cardy
condition \cite{cardy};
\begin{equation}
 \cA^i _{ab} = \sum_j B^j _a B^j_b S^{~i}_{j} ~.
\label{C_cond}
\end{equation}  

The analysis of M\"{o}bius strip amplitudes is a little more complicated 
than the other cases.
It can be seen that the following characters are convenient to study the 
M\"{o}bius strip amplitudes
\begin{equation}
 \hat{\chi}_j (q) \equiv e^{-  \pi i (h_j - \frac{c}{24})} 
 \chi_j (- \sqrt{q})~.
\label{mobius}
\end{equation}
The transformation from the direct channel to the transverse channel can
be obtained
by the $P (\equiv \sqrt{T} S T^2 S \sqrt{T}$)\footnote{
$\sqrt{T}$ is defined by the square root of the phase shift under the
$T$ transformation.
} 
matrix \cite{sagnotti1}.
In $SU(2)$ WZW model, this matrix can be written as
\begin{equation}
 P_{j j'}^{SU(2)} = \frac{2}{\sqrt{k+2}} 
   \sin \left( \frac{\pi (2 j + 1)(2 j' + 1)}{2(k+2)}\right) 
  E_{k+2j+2j'}~,
\end{equation}
where we use
\begin{equation}
 E_{n} = \frac{1+(-1)^{n}}{2} ~.
\label{En}
\end{equation}
The modulus of M\"{o}bius strip can be assigned as $\tilde{\tau} = 2 i t$
and $P$ transforms $t \to 1/(4t)$. 
The M\"{o}bius strip amplitudes in the direct channel are given by
\begin{equation}
 \cM _a (q) = \tr _{\cH_a}(\Omega h q^{L_0 - \frac{c}{24}}) 
              = \sum_i \cM_a^i \hat{\chi}_i(q) ~,
\end{equation}
where there is only one boundary condition $a$ and
the coefficients $\cM_a^i$ are the integers.
The amplitudes in the transverse channel are obtained by the overlaps
between the crosscap states and the boundary states
\begin{equation}
 \cM (\tq) = ~_C \bra{C} \tq^{H_c} \ket{B,a}_C 
         = \sum_i \Gamma^i B^i_a \hat{\chi}_i (\tq)~.
\end{equation}
As we said above, the modular transformation is given by $P$
transformation, thus the constraint becomes
\begin{equation}
 \cM^i_a = \sum_j  \Gamma^j B^j_a P_{j}^{~i} ~.
\end{equation}
The presence of $P$ matrix in this equation makes things different from
the case of the annulus amplitudes.

\begin{figure}
\centerline{\includegraphics{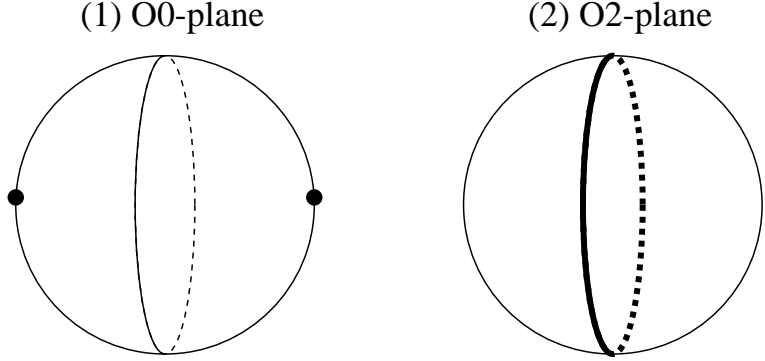}}
\caption{The geometry of orientifolds of $SU(2)$ WZW models.
 (1) The crosscap state of the type (\ref{BS}) describes two
 $\cO$0-planes located at the opposite points of $S^3$. 
 (2) The  crosscap state of the type (\ref{BS2}) describes 
 $\cO$2-plane whose geometry is $S^2$ at the equator of $S^3$.
}
\label{su2oplane}
\end{figure}

Moreover we have to care about 
so called completeness conditions \cite{sagnotti2}.
The following set of crosscap states and boundary states
which obey all consistency conditions was obtained  
\cite{sagnotti1,sagnotti2,sagnotti3,sagnotti4} by the coefficients
\begin{equation}
 B^{i}_a = \frac{S_{ai}}{\sqrt{S_{0i}}}
~,~~\Gamma^i = \frac{P_{0i}}{\sqrt{S_{0i}}} ~.
\label{BS}
\end{equation}
If we use the Verlinde formula of the fusion coefficients 
\begin{equation}
 N_{i,j}^{~~l} \equiv \sum_m \frac{S_{mi} S_{mj} S_{m}^{~l}}{S_{m0}}~,
\end{equation}
and define the integer valued tensor as \cite{sagnotti1}
\begin{equation}
 Y_{i,j}^{~~l} \equiv \sum_m \frac{S_{mi} P_{mj} P_{m}^{~l}}{S_{m0}}~,
\end{equation}
the amplitudes can be obtained as
\begin{equation}
 \cK (q) = \sum_i Y^i_{~0,0} \chi_i (q) ~,~~
 \cA_{ab} (q) = \sum_i N^{~~i}_{a,b} \chi _i (q) ~,~~
 \cM_a (q) = \sum_i Y_{a,0}^{~~i} \hat{\chi}_i (q) ~.
\end{equation}
Let us see the example of $SU(2)_k$ WZW models.
The $Y$ matrices are given as
\begin{equation}
 Y^i_{~0,0} = (-1)^{2i} ~, ~~Y_{a,0}^{~~i} = (-1)^{2a} (-1)^i N^{~~i}_{a,a} ~,
\label{Yi00}
\end{equation}
and the fusion matrices are 
\begin{equation}
 N_{i, j }^{~~l } = \left\{
\begin{array}{rcl}
1 && | i - j | \leq l \leq \mbox{min} \{ i +j , k - i - j\}~,~~
       i + j  + l  \in \bz \\
0 && \mbox{otherwise} ~.
\end{array} 
\right.
\end{equation}  
The orientifolds in this model were investigated  geometrically in 
\cite{oplane1,oplane3,oplane3} and this type of
orientifold is identified as $\cO$0-planes, see figure 
\ref{su2oplane}.  
 
The other kind of orientifolds are elegantly expressed by using the
simple current technique \cite{simple1,simple2}.
The simple currents are the fields $J$ which satisfy\footnote{
We only analyze level 2 simple currents, namely, $J \times J = 1$.
The level $N$ simple currents can be used to construct the crosscap states 
in $SU(N)$ WZW models \cite{simple1}.}
\begin{equation}
 J \times i = i' ~(~^\forall i)~.
\end{equation}
The trivial simple current is identity $0$ and then $i' = i$.
For $SU(2)_k$ WZW model, there is a non-trivial simple current
$J=k/2$ and then $i' = k/2 - i$. 
By applying simple currents to the previous results, the set of crosscap
states and boundary states which satisfy all consistency conditions was
obtained by the coefficients 
\begin{equation}
 B^i_a = \frac{S_{ai}}{\sqrt{S_{Ji}}}
~,~~\Gamma^i = \frac{P_{Ji}}{\sqrt{S_{Ji}}} ~.
\label{BS2}
\end{equation}
We have changed also boundary states in the above equation for convenience,
however these states are not new but the renamed ones.
The amplitudes can be written as
\begin{equation}
 \cK (q) = \sum_{~~i} Y^i_{J,J} \chi_i (q) ~,~~
 \cA_{ab} (q) = \sum_i N^{~~J \times i}_{a,b} \chi _i (q) ~,~~
 \cM_a (q) = \sum_i Y_{a,J}^{~~i} \hat{\chi}_i (q) ~.
\end{equation}
In $SU(2)_k$ WZW  models, these $Y$ matrices are given as
\begin{equation}
 Y^i_{~\frac{k}{2},\frac{k}{2} } = 1 ~,~~
 Y_{a, \frac{k}{2}} ^{~~i} = N^{~~\frac{k}{2}-i}_{a,a} ~,
\end{equation}
and this type of orientifold corresponds to $\cO$2-plane at the equator
of the $S^3$ \cite{oplane1,oplane3,oplane3}, see figure \ref{su2oplane}.

%%%%%%%%%%%%%%%%%%%%%%%%%%%%%%%%%%%%%%%%%%%%%%%%%%%%%%%%%%%%%%%%%%%%%

\section{$U(1)$ WZW Models and Orientifolds}
\indent

In this section, we study the orientifolds of $U(1)_k$ WZW model.
It can be seen that the $\bz_k$ orbifold of this theory is T-dual to the
original theory, 
therefore we can apply the construction proposed by \cite{MMS} to the
orientifolds.
We shall show that 
the orientifolds which correspond to the solution (\ref{BS}) constructed
in the previous section are $\cO$0-planes.
Then we construct the B-type orientifold by applying  $\bz_k$ orbifold
and T-duality and it can be identified as $\cO$1-plane wrapping the whole
line. 
In the next subsection, we investigate the closed strings and D-branes
in $U(1)$ WZW models, and in subsection 3.2, we construct the orientifolds.
 
\subsection{$U(1)$ WZW Models and D-branes}
\indent

The $U(1)_k$ WZW model has the current $J = i \sqrt{2 k } \partial X$
and the 
primaries $\Phi_n (X)  = \exp (i \frac{n}{\sqrt{2k}} X)$ whose
labels represent their $U(1)$ charges and their conformal
weights are $h_n = n^2/(4k)$.
The label $n$ is defined modulo $2k$ and we take $-k+1, -k+2, \cdots,
k$. The characters are defined as
\begin{equation}
 \psi_n (q,z) = \tr _{\cH_n} q^{L_0 - 1/24} e^{2 \pi i z J_0} = 
 \frac{\Theta_{n,k} (\tau,2z) }{\eta(\tau)} ~,
\end{equation} 
and the characters with $z=0$ will be used 
\begin{equation}
 \psi _n ( q) = 
  \frac{1}{\eta (q)} \sum_{r \in \bz} q^{k(r + \frac{n}{2k})^2}~.
\label{U1char}
\end{equation}
Their modular transformations are given by
 \begin{equation}
 \psi_n (\tau + 1) =
  e^{2 \pi i (h_n - 1/24)} \psi_n (\tau) ~,~~
 \psi_n (- \frac{1}{\tau}) =
  \frac{1}{\sqrt{2k}} \sum_{n'} e^{- \pi i n n'/ k } \psi_{n'}(\tau) ~,
\end{equation} 
and the diagonal modular invariant is given by
\begin{equation}
 \cT = \sum_n |\psi_n (q)|^2 ~.
\end{equation} 
This theory is the same as the one free boson theory with radius $\sqrt{2k}$
where $\alpha ' = 2$.  
This theory has $\bz_k$ global symmetry under the transformation
\begin{equation}
 g: \Phi_n \to e^{2 \pi i n /k} \Phi_n ~,
\end{equation}
and we can make the orbifold by this transformation. 
The most general modular invariants are given by $\bz_l$ orbifold
procedure $(k = ll', l,l' \in \bz)$ which are generated by $g^{l'}$ 
\begin{equation}
\cT  = 
 \sum_{n,\bar{n}}
 \psi_n (q) \psi_{\bar{n}} (\bar{q}) ~,~~
n+\bar{n}=0 \mod 2l ~,~ n-\bar{n}=0 \mod 2l' ~.
\end{equation} 
From these modular invariant combinations, 
we can see that  
the original $U(1)_k$ WZW model is T-dual to its $\bz
_k$ orbifold model since $\psi_n = \psi_{-n}$.

It is well-known that there are two types of D-branes in the $U(1)$ WZW
model. The open strings on the one type of D-branes obey Neumann
boundary condition and the ones on the other obey Dirichlet boundary
condition. 
We will call the Dirichlet condition as A-type and the Neumann one as
B-type. 
The D-branes are represented in the terms of the boundary states.    
The general boundary states which satisfy A-type boundary condition 
$J(z) + \tilde{J}(\tilde{z}) =0$\footnote{The boundary condition is
assigned in the closed string channel and we use the notation $J = i
\sqrt{2k} \partial X$ and $\tilde{J} = - i \sqrt{2k} \tilde{\partial} X$. }  
are given by Ishibashi states \cite{ishibashi} 
\begin{equation}
 \ket{A,r,r}_I = 
  \exp\left(\sum_{n=1}^{\infty} \frac{1}{n} \alpha_{-n} \tilde{\alpha}_{-n}
       \right)  
     \sum_{l \in \bz} 
     \kett{ \frac{r + 2 kl}{\sqrt{2k}}, \frac{r + 2 kl}{\sqrt{2k}}} ~.
\end{equation}   
The label $r$ runs $-k+1, -k+2, \cdots, k$. The boundary states
corresponding to D-branes have to satisfy Cardy conditions
(\ref{C_cond}) and they are obtained by (\ref{BS})
\begin{equation}
 \ket{A,\hat{n}}_C = \sum_n \frac{S_{\hat{n}n}}{\sqrt{S_{0n}}}
 \ket{A,n,n}_I 
  = \frac{1}{(2k)^{1/4}} \sum_n e^{- \pi i \hat{n} n /k} 
 \ket{A,n,n}_I ~.
\end{equation}
These states describe D0-branes and their locations are represented by
the label $\hat{n} = -k+1, -k+2, \cdots, k$.
The $\bz_k$ symmetry corresponds to the rotation of the circle and $g$
transforms $\ket{A,\hat{n}}$ to $\ket{A,\hat{n} - 2}$. 
The annulus amplitudes between A-type branes are given by 
\begin{equation}
  \cA = ~_C \bra{A,\hat{n}} \tq^{H_c} \ket{A, \hat{n}'}_C 
   = \psi_{\hat{n}- \hat{n}'} (q) ~.
\end{equation}
In non-Abelian WZW models, A-branes correspond to the maximally symmetric
branes which have been widely investigated.

Neumann boundary condition is described by the currents as 
$J(z) - \tilde{J}(\tilde{z}) = 0$ and B-type Ishibashi states are
given by
\begin{equation}
 \ket{B,r,-r}_I = \exp \left( - \sum_{n=1}^{\infty} 
  \frac{1}{n}\alpha_{-n} \tilde{\alpha}_{-n} \right) 
     \sum_{l \in \bz} 
     \kett{ \frac{r + 2 kl}{\sqrt{2k}},- \frac{r + 2 kl}{\sqrt{2k}}} ~.
\label{Br-r}
\end{equation} 
We should note here that there are only $r=0,k$ states in the bulk
spectrum. 
The Cardy states in this case are given by
\begin{equation}
 \ket{B, \eta = \pm 1}_C = \left(\frac{k}{2}\right)^{1/4} 
\left( \ket{B,0,0}_I + \eta \ket{B,k,-k}_I \right) ~, 
\end{equation}
where the coefficients are chosen to have the consistent open spectrum.
These boundary states describe the D1-branes and $\eta$ represents the
Wilson line.  
These states can be also obtained by using $\bz_k$ orbifold and
T-duality \cite{MMS}. 
To get the boundary states in this orbifold theory, we sum over all
image branes generated by $g^m $ $(m = 0,1,\cdots,k-1)$.
The amplitudes between B-type branes are obtained as  
\begin{equation}
  \cA  = ~_C \bra{B,\eta} \tq^{H_c} \ket{B, \eta'}_C 
   = \sum_m \frac{1 + \eta \eta' (-1)^m}{2} \psi_m(q) ~.
\end{equation}

The amplitudes between A-type and B-type branes have the different
open string spectrum.
The amplitudes between Ishibashi states are given by 
\begin{equation}
 ~_I \bra{A,n,n} \tq ^{H_c} \ket{B,r,-r}_I = 
  \delta_{n,0} \delta_{r,0} \chi_{ND} (\tq) ~,
\end{equation} 
where we defined $\chi_{ND}$ as
\begin{equation}
 \chi_{ND} (\tq) = \frac{1}{\tq^{1/24} \prod_{n=1}^{\infty} (1+ \tq^n)} ~,
\label{chiND}
\end{equation}
and its modular transformation is given by
\begin{equation}
\chi_{ND} (\tq) = \sqrt{2} \chi_{ND}' (q) ~,~~
 \chi_{ND}' (q) = 
  \frac{q^{\frac{1}{48}}}{ \prod_{n=1}^{\infty} (1 - q^{n-\frac{1}{2}})}
 ~.
\end{equation}
Using these characters, the annulus amplitudes are obtained as
\begin{equation}
 \cA = ~_C \bra{A,\hat{n}} \tq^{H_c} \ket{B,\eta}_C
  = \chi ' _{ND} (q) ~.
\end{equation}
It can be seen that the character $\chi'_{ND}$ correctly reproduces the
spectrum of open strings which satisfy Neumann-Dirichlet boundary condition.

\subsection{Crosscap States in $U(1)$ WZW Models}
\indent

Next we shall construct the crosscap states which represent the
orientifolds. The most famous example is Type I string theory, which can
be regarded as the system with $\cO$9-planes of Type IIB string
theory. Applying  $(9-p)$ T-dualities to $\cO$9-planes, we have 
$\cO p$-planes.  
Therefore we have $\cO$0-planes and $\cO$1-planes in our $U(1)$
WZW models.   

To construct the crosscap states, we have to obtain $P$ matrix in
$U(1)$ WZW model as explained in the previous section.
It is useful to calculate the next quantity 
by using the modular transformation of $U(1)$ characters as
\begin{eqnarray}
 (ST^2 S)_{mn} &=& \frac{1}{2k} \sum_{l= -k +1} ^{k}
  e^{- \pi i m l /k} 
  e^{4 \pi i \left( \frac{l^2}{4k} -  \frac{1}{24}\right)}
 e^{- \pi i l n/k} \nn
  &=& \frac{1}{2k} \sum_{l= -k+1} ^{k} 
   e^{ \frac{\pi i}{k}\left(l -  \frac{m+n}{2}\right)^2}
   e^{- \frac{\pi i}{4k} (m+n)^2 - \frac{\pi i}{6}} ~.
  \end{eqnarray} 
Using the formula of Gaussian sum 
\begin{eqnarray}
 \sum_{n=0}^{\kappa -1} e^{2 \pi i n^2 / \kappa } &=& 
\frac{1}{2} (1+i)(1+i^{- \kappa }) \sqrt{ \kappa} \nn
 \sum_{n=0}^{\kappa -1} e^{2 \pi i (n + 1/2)^2 / \kappa} &=& 
\frac{1}{2} (1+i)(1-i^{- \kappa }) \sqrt{\kappa} ~,
\end{eqnarray}
we can obtain the $P$ matrix for $U(1)$ WZW models as
\begin{equation}
 P_{mn} = \frac{1}{\sqrt{k}} e^{- \pi i m n /2k} E_{k+m+n} ~,
\label{U1P}
\end{equation}
where $E_{k+m+n}$ is the projection to the even elements (\ref{En}).

In the presence of orientifolds, 
we have to add some conditions to the right and left moving currents
just like the boundary conditions.
In our case, these conditions are written as
\begin{equation}
 (J_n \pm (-1)^n \tilde{J}_{-n}) \ket{O} = 0~.
\label{CC}
\end{equation}
We call the crosscap states obeying the $(+)$ condition as A-type
and the ones obeying the $(-)$ condition as B-type.
The Ishibashi crosscap states can be easily constructed and A-type ones are 
\begin{equation}
 \ket{OA,r,r}_I = \exp\left(\sum_{n=1}^{\infty} \frac{(-1)^n}{n}
   \alpha_{-n} \tilde{\alpha}_{-n} \right)  \sum_{l \in \bz} 
     \kett{ \frac{r + 2 kl}{\sqrt{2k}}, \frac{r + 2 kl}{\sqrt{2k}}} ~,
\end{equation}   
where the label $r$ runs $-k+1, -k+2, \cdots, k$. 
This type of crosscap states are the ones discussed in the previous
section and the Cardy crosscap states are given by (\ref{BS}) 
\begin{equation}
 \ket{OA,\hat{n}}_C = \sum_n \frac{P_{\hat{n}n}}{\sqrt{S_{0n}}}
 \ket{OA,n,n}_I ~. 
\end{equation}
The Klein bottle amplitudes can be obtained by using the crosscap states
as
\begin{equation}
 \cA = ~_C \bra{OA,\hat{n}} \tq^{H_c} \ket{OA,\hat{n}} _C
    = \psi_0 (q) + (-1)^{\hat{n}+k} \psi_k (q) ~. 
\label{U1AKB}
\end{equation}
We can see that this is the appropriate orientifold projection because
there is the summation of the ``winding number'' $l$ in the character
(\ref{U1char}).
From this reason, this crosscap state can be identified as
$\cO$0-plane.   

The M\"{o}bius strip amplitudes can be given by the overlaps between the
boundary states and the crosscap states. If we use A-type boundary states, we
obtain 
\begin{equation}
 \cM  = ~_C \bra{A,\hat{n}} \tq^{H_c} \ket{OA,\hat{n}'} _C
   = \hat{\psi}_{2 \hat{n} - \hat{n}'} (q) ~.
\end{equation}
These amplitudes show that the positions of orientifolds are the same
points of D-branes with the label $\hat{n}=\hat{n}'/2, \hat{n}'/2 + k$ in the
even $\hat{n}'$ case.
The odd  $\hat{n}'$ orientifolds are located at the middle point between 
$\hat{n} = (\hat{n}' -1 )/2$ and   $\hat{n} = (\hat{n}' + 1 )/2$
and the opposite points of the circle.
These points correspond to the fixed points under the reflection. 
In fact, the amplitudes with mirror branes are 
\begin{equation}
 \cA  
= ~_C \bra{A,\hat{n}} \tq^{H_c} \ket{A,\hat{n}' - \hat{n}} _C
   = \psi_{2 \hat{n} - \hat{n}'} (q) ~,
\end{equation}
whose spectrum is the same as in the M\"{o}bius strip amplitudes.

The calculation of the amplitudes with B-type boundary states needs $P$
transformation of the character $\chi_{ND}$ (\ref{chiND}).
Using the fact that this character can be rewritten as
\begin{equation}
 \chi_{ND} (q) = \frac{1}{q^{1/24} \prod_{n=1}^{\infty} (1+ q^n)} 
   = \frac{\eta(\tau)}{\eta(2\tau)} ~,
\end{equation}
and the modular transformation of $\eta$ functions (\ref{etam}), 
we get 
\begin{equation}
 \hat{\chi}_{ND} (\tq) = \hat{\chi}_{ND} (q) ~,
 \label{PND}
\end{equation}
namely, $P=1$. 
Hence we can get the amplitudes as
\begin{equation}
 \cM  = ~_C \bra{B,\eta} \tq^{H_c} \ket{OA,\hat{n}} _C
   = E_{\hat{n}+k} \hat{\chi}_{ND} (q) ~.
\end{equation}
This is the spectrum of the open strings between D1-brane in the system with 
$\cO$0-plane.

\begin{figure}
\centerline{\includegraphics{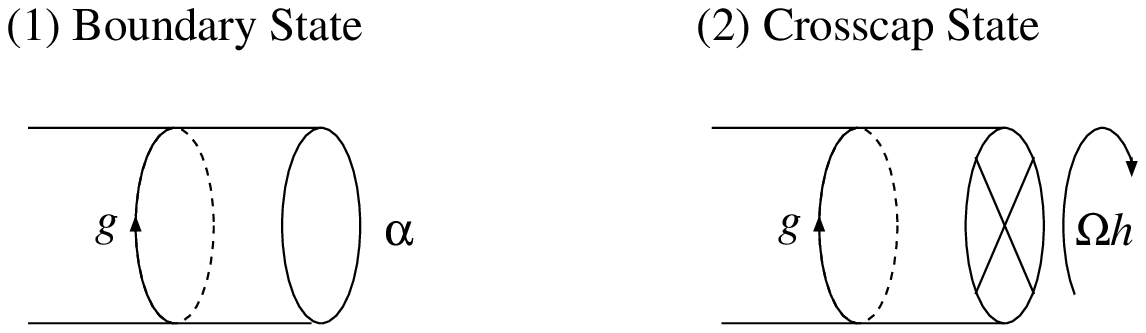}}
\caption{Boundary states and crosscap states in $g$ twisted sector. 
 (1) Boundary conditions are denoted by $\alpha$, which must be invariant
 under the operation $g$. (2) Opposite points of crosscap states must be
 related by $\Omega h$, where $(\Omega h)^2 = g$.  
}
\label{twist}
\end{figure}

The B-type orientifolds can be constructed in the way similar to the B-type
branes. 
The Ishibashi crosscap states obeying the condition (\ref{CC}) with
$(-)$ are given by
\begin{equation}
 \ket{OB,r,-r}_I = \exp \left( - \sum_{n=1}^{\infty} 
  \frac{(-1)^n}{n}\alpha_{-n} \tilde{\alpha}_{-n} \right) 
     \sum_{l \in \bz} 
     \kett{ \frac{r + 2 kl}{\sqrt{2k}},- \frac{r + 2 kl}{\sqrt{2k}}} ~,
\end{equation} 
with $r=0,k$, which is the same as the B-brane case. 
The Cardy crosscap states in this case can be obtained by using
$Z_{k}$ orbifold and T-duality.
In $g$ twisted sector, the orientifold operation $\Omega h$  must be
$(\Omega h)^2 = g$ \cite{GP}, see figure \ref{twist}.
$g^{1/2}$ generates the half sift of $g$ action to the branes and the states 
$\ket{OA, \hat{n}}$ are transformed to the states $\ket{OA, \hat{n}-2}$.
Summing over all image A-type crosscap states, 
we get  (for even $k$ case)
\begin{equation}
 \frac{1}{\sqrt{k}} \sum_{\hat{m}=0}^{k-1} \ket{OA,2 \hat{m}}_C 
 = (2k)^{1/4} \ket{OA,0,0}_I  ~.
\label{U1orbifold}
\end{equation}
In odd $k$ case, we can use the odd $\hat{n}$ crosscap states instead.
Applying T-duality, we can get the B-type crosscap state as 
\begin{equation}
 \ket{OB}_C 
 = (2k)^{1/4}  \ket{OB,0,0}_I ~, 
\end{equation}
which is the same as the usual crosscap states for $\cO$1-plane.
In fact, the Klein bottle amplitude is given by
\begin{equation}
 \cA = ~_C \bra{OB} \tq^{H_c} \ket{OB} _C
    = \sum_n \psi_n (q) ~,
\end{equation}
which is the summation of the states with all momentum.
The M\"{o}bius strip amplitudes with B-type boundary states can be
obtained as
\begin{equation}
 \cM  = ~_C \bra{B,\eta} \tq^{H_c} \ket{OB} _C
   = \sum_n E_{n+k} \hat{\psi}_{n} (q) ~,
\end{equation}
and the amplitudes with A-type boundary states are given by
\begin{equation}
 \cM  = ~_C \bra{A,\hat{n}} \tq^{H_c} \ket{OB} _C
   = \hat{\chi}_{ND} (q) ~,
\end{equation}
where we use the $P$ transformation of $\hat{\chi}_{ND}$ (\ref{PND}).

%%%%%%%%%%%%%%%%%%%%%%%%%%%%%%%%%%%%%%%%%%%%%%%%%%%%%%%%%%%%%%%%%%%

\section{$SU(2)/U(1)$ WZW Models and Orientifolds}
\indent

In this section, we construct the orientifolds of $SU(2)_k/U(1)_k$ WZW
model. 
The A-type orientifolds can be obtained by Cardy construction (\ref{BS})
and they are identified as $\cO$1-planes connecting the opposite
boundary points.
Applying the $\bz_k$ orbifold and T-duality, we can get two types of
B-type orientifolds which are $\cO$0-plane at the center of
the disk and $\cO$2-plane wrapping the whole disk.
In the next subsection, we study the closed strings in the 
parafermionic theory and construct D-branes followed by \cite{MMS}.
In subsection 4.2, the crosscap states are
constructed.    

\subsection{$SU(2)/U(1)$ WZW Models and D-branes}
\indent
   
The parafermionic theory can be described by the $SU(2)_k/U(1)_k$ coset WZW
model \cite{GQ}, which is 
defined by gauging $U(1)$ sector in the 
$SU(2)$ WZW model. We denote the Hilbert space of  $SU(2)_k$ WZW model
as $\cH^{SU(2)}_{j,m}$, where $j$ is the Casimir number
and $m$ is the $J^3_0$ eigenvalue of the highest weight
representation. These values take $j = 0, 1/2, \cdots,
k/2$ and $ m = -2j, -2j+1,\cdots, 2j $. 
We also denote the parafermion Hilbert space as $\cH^{PF}_{j,n}$,
where $j =  0, 1/2, \cdots, k/2$ and $ n \in \bz_{2k} $
and $U(1)$ Hilbert space of momentum $n$ is denoted as $\cH^{U(1)}_n$.
These Hilbert spaces are related as
\begin{equation}
 \cH^{SU(2)}_{j,m} = \cH^{PF}_{j,m} \otimes \cH^{U(1)}_m ~.
\end{equation}
The right hand side does not include all Hilbert space, in fact, the
parafermionic Hilbert space has the following identification
\begin{equation}
 \cH^{PF}_{j,n} = \cH^{PF}_{k/2 - j, n+k} ~,
\label{PFI}
\end{equation} 
which will be often called as the ``spectral flow'' identification. 
There is also the relation between $j$ and $n$ as $2j+n
\in 2 \bz$.
From now on, we will call the distinct irreducible representations as
$PF(k)$. 

The characters of parafermionic theory can be given by the
characters of $SU(2)$ and $U(1)$ currents as
\begin{equation}
 \chi_j ^{SU(2)} (\tau,z) = 
  \sum_{n=-k+1} ^{k} \chi_{j,n} (\tau) \psi_n (\tau,z) ~,
\end{equation}
which have the following identities as
\begin{equation}
 \chi_{j,n}(q) = \chi_{j,-n}(q) = \chi_{k/2 - j, k-n}(q) ~.
\label{char-rel}
\end{equation}
The lowest conformal weights of $PF(k)$ are given by
\begin{equation}
 h_{j,n} = \left\{
\begin{array}{ll}
 \frac{j(j+1)}{k+2} - \frac{n^2}{4k} & -2j \leq n \leq 2j \\
 \frac{j(j+1)}{k+2} - \frac{n^2}{4k} + \frac{n-2j}{2} &
 2j \leq n \leq 2k-2j ~,
\end{array}
\right. 
\label{pfweight}
\end{equation} 
where we use $n = -2j, -2j+2, \cdots, 2k - 2j - 2$.
The modular transformations of the parafermionic characters are given by
\begin{eqnarray}
 \chi_{j,n} (\tau +1) &=& e^{2 \pi i ( h_{j,n} - \frac{c}{24})}
 \chi_{j,n} (\tau)  \nn
 \chi_{j,n} (- 1/ \tau) &=&
  \sum_{(j',n') \in PF(k)} S_{j,n}^{PF~j',n'} \chi_{j', n'} (\tau) ~,
\end{eqnarray}
where we use
\begin{equation}
  S_{j,n}^{PF~j',n'} =\sqrt{\frac{2}{k}}
      {S^{SU(2)}}_j^{~j'} e^{\pi i n n'/k} ~.
\end{equation}
The fusion coefficients can be written as
\begin{equation}
 N_{(j,n),(j',n')}^{PF~~(j^{''},n^{''})} 
 = N_{j,j'}^{~~j^{''}} \delta^{(2k)}_{n+n'-n^{''}} +
  N_{j,j'}^{~~k/2 - j^{''}} \delta^{(2k)}_{n+n'-n^{''} -k} ~,
\end{equation}
where we define $\delta^{(2k)}$ modulo $2k$.

The diagonal modular invariant is given by
\begin{equation}
 \cT = \sum_{(j,n) \in PF(k)} |\chi_{j,n} (q)|^2 ~.
\end{equation}
Moreover the orbifold procedure under the global $\bz_k$ symmetry
\begin{equation}
 g: \Phi_{j,n} \to e^{2 \pi i n/k} \Phi_{j,n}
\end{equation}
gives other modular invariants. 
The torus amplitudes of $\bz_l$ orbifolds ($k=l l'$) are given by
\begin{equation}
 \cT = \frac{1}{2} \sum_{j,n,\bar{n}} \chi_{j,n} (q) 
 \bar{\chi}_{j,\bar{n}} (\bar{q}) 
 ~,~ n-\bar{n} =0 \mod 2l ~,~ n+\bar{n} =0 \mod 2l' ~. 
\end{equation}    
Just as the $U(1)$ case, the $\bz_k$ orbifold and T-duality gives the same
spectrum of the original theory 
because of the character identities (\ref{char-rel}).

\begin{figure}
\centerline{\includegraphics{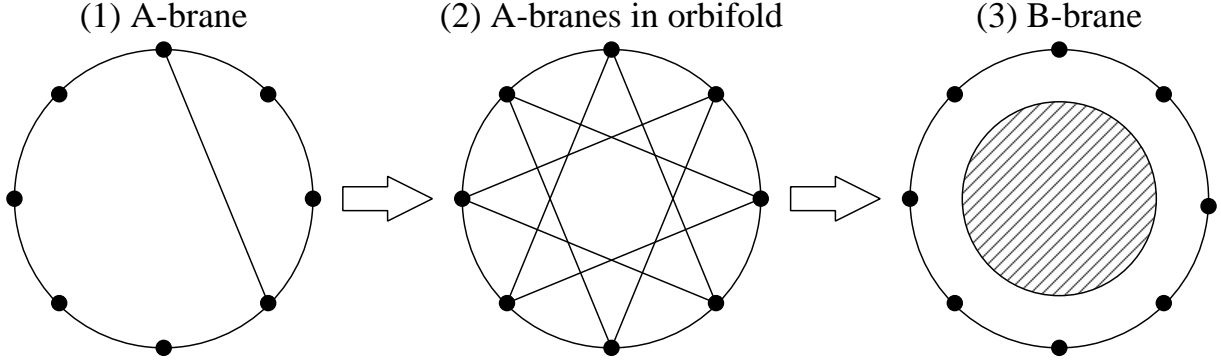}}
\caption{The geometry of D-branes in $SU(2)_k/U(1)_k$ WZW model with $k=8$.
 The line in (1) describes A-type D1-brane with $\hat{j}= 3/2$. 
 The summation of image branes is given in (2). 
 Applying T-duality, we obtain B-brane with $\hat{j}= 3/2$ as (3).  
}
\label{branes}
\end{figure}

Next we study D-branes in the parafermionic theory.
It was found in \cite{MMS} that there are two types of branes, which
are called as A-type and B-type branes.
The A-type branes correspond to the solution (\ref{BS}).
Using A-type Ishibashi states $\ket{A,j,n}_I$, they can be
constructed as  
\begin{equation}
 \ket{A,\hat{j}, \hat{n}}_C = \sum_{(j,n)\in PF(k)} 
  \frac{S_{\hat{j},\hat{n}}^{PF~j,n}}{\sqrt{S^{PF~j,n}_{0,0}}}
  \ket{A,j,n}_I ~.
\end{equation}
The annulus amplitudes between these A-type boundary states can be
calculated by using the Verlinde formula as
\begin{equation}
 ~_C \bra{A,\hat{j},\hat{n}} \tq^{H_c} \ket{A,\hat{j}',\hat{n}'}_C
  = \sum_{(j,n) \in PF(k)} 
   N^{PF~~~~~(j,n)}_{(\hat{j},-\hat{n}),(\hat{j}',\hat{n}')} 
 \chi_{j,n} (q)~.
\end{equation} 
The target space is given by the disk with radius one (see Appendix B)
and the geometry of
the branes can be determined by scattering with the closed string states
\cite{MMS}.
The boundary states with $\hat{j}=0$ describe D0-branes at $k$ special
points of the boundary of the disk and 
the general $\hat{j}$ states connect $2 \hat{j}$ separated points as in
figure \ref{branes}. The rotation of the disk gives the states with different
$\hat{n}$. 

The B-type branes can be constructed by using $\bz_k$ orbifold procedure and
T-duality from A-type branes. 
The $\bz_k$ orbifold projects to the states with $n=0,k$ and we will
only use $n=0$ states by making use of the spectral flow 
identification (\ref{PFI}).
The T-duality in parafermionic theory can be realized by the operator
$\exp(\pi i J^1_0)$.
We can see that this operator changes $\Phi_{j,n}$ into $\Phi_{j,-n}$
because of the property 
\begin{equation}
  e^{\pi i J^1_0} J^3_0 e^{- \pi i J^1_0} = - J^3_0 ~. 
\end{equation}
In this way, we can define the B-type Ishibashi crosscap states as
\begin{equation}
 (1 \otimes e^{\pi i \tilde{J}^1_0}) \ket{j}^{SU(2)}_I 
  = \sum^{k}_{r=-k+1} \ket{B,j,r,-r}_I^{PF} \otimes \ket{B,r,-r}_I^{U(1)} ~, 
\label{BPFI}
\end{equation}
where $\ket{B,r,-r}^{U(1)}$ are defined in (\ref{Br-r}).
By combining the $\bz_k$ orbifold procedure and T-duality, the Cardy
states can be constructed from these Ishibashi states as 
\begin{equation}
 \ket{B,\hat{j}}_C = 
  (2k)^{1/4} \sum_{j \in \bz} \frac{S_{\hat{j}}^{~j}}{\sqrt{S_0^{~j}}}
  \ket{B,j,0,0}_I ~,
\label{B-brane}
\end{equation}
and the annulus amplitudes between two B-branes are calculated as
\begin{equation}
 A  = ~_C \bra{B,\hat{j}} \tq^{H_c} \ket{B,\hat{j}'}_C 
  = \sum_{j = 0}^{k/2} \sum_{n =- 2 j }^{2k -2 j  -2} 
     N_{\hat{j},\hat{j}'}^{~~j} \chi_{j ,n } (q) ~. 
\end{equation}
The geometry of B-branes is given in figure \ref{branes}.
From A-branes we can construct B-branes by summing
all image branes and taking T-duality.
Therefore we can see that the state with $\hat{j}=0$ corresponds to
D0-brane at the center of the disk and the states with the general $\hat{j}$
correspond to D2-branes.    
In even $k$ case, we can construct $\hat{j}=k/4$ B-brane and there are two
identity states in the open string spectrum between these branes.  
Therefore we need to redefine this brane to be the irreducible one,
however we do not deal with this brane for simplicity  and
analyze only general $\hat{j} \neq k/4$ B-branes. 

The annulus amplitudes between A-type and B-type branes are 
obtained by using the overlaps between Ishibashi states 
\begin{equation}
 ~_I \bra{A,j,0,0} q^{H_c} \ket{B,j',0,0} _I 
 = \delta_{j,j'} \chi'_j (q) ~,
\end{equation}
where 
\begin{equation}
 \chi'_j (q) = \frac{\chi^{SU(2)}_j (q,z=1/2)}{\chi_{ND} (q)} ~.
\label{chi'}
\end{equation}
Their modular transformations are given by
\begin{equation}
 \chi'_j(\tq) = \sum_{j'=0}^{k/2}
 \frac{1}{\sqrt{2}} S_{j}^{~j'} \tilde{\chi}_{j'} (q) ~,
\end{equation}
where we define
\begin{equation}
 \tilde{\chi}_j (q) = e^{\pi i k \tau /8} \chi_j ^{SU(2)} (\tau, \tau/2) 
                      q^{-1/48}\prod_{n=1}^{\infty} (1-q^{n-1/2})  ~.
\end{equation}
Now we can get the amplitudes as
\begin{equation}
\cA =  ~_C \bra{B, \hat{j}} \tq^{H_c} \ket{A,\hat{j}',\hat{n}'}_C
 = \sum_{j=0}^{k/2} 
  N_{\hat{j},\hat{j}'}^{~~j} \tilde{\chi}_{j} (q) ~. 
\end{equation}

\subsection{Crosscap States in $SU(2)/U(1)$ WZW Models}
\indent

Before constructing the crosscap states, we have to know the explicit
form of $P$ matrix.
Since there is the spectral flow identification (\ref{PFI}), 
we have to be careful to make the $P$ matrix. We find 
\begin{eqnarray}
 \hat{\chi}_{j,n}(q) 
   &=& \frac{1}{2}    \sum_{j'=0}^{k/2} \sum_{n' = -2j'}^{2k -2j' -2}
 \left(  P_{j}^{~j'} P_{n}^{~n'} \hat{\chi}_{j',n'} (q)
  + P_{j}^{~k/2-j'} P_{n}^{~n'+k} 
  \hat{\chi}_{k/2 - j',n' + k} (q) \right) \nn
  &=& \frac{1}{2}    \sum_{j'=0}^{k/2} \sum_{n' = -2j'}^{2k -2j' -2}
 \left(  P_{j}^{~j'} P_{n}^{~n'}  + (-1)^{j' - n'/2}
  P_{j}^{~k/2-j'} P_{n}^{~n'+k} \right)
  \hat{\chi}_{j',n' } (q)  ~,
\end{eqnarray}
where we use $P_j^{~j'}$ for $SU(2)$ $P$ matrix and $P_n^{~n'}$
for the inverse of $U(1)$ $P$ matrix
\begin{equation}
P_n^{~n'} =  \frac{1}{\sqrt{k}} e^{\pi i n n' /2k} E_{n + n' + k} ~.
\end{equation}
In the last line, we used that   
\begin{equation}
 \hat{\chi}_{k/2-j,n+k} (q) = (-1)^{j - n/2} \hat{\chi}_{j , n} (q) ~,
\label{char-SF}
\end{equation}
since we have defined the character as (\ref{mobius}) and the conformal
weights are given by (\ref{pfweight}).
Therefore the $P$ matrix which is consistent with the spectral flow
identification (\ref{PFI}) is given by
\begin{equation}
 P^{PF ~ j',n'}_{j,n} =  
  P_{j}^{~j'} P_n^{~n'}  
  + (-1)^{j' - n'/2} P_{j}^{~k/2-j'} P_n^{~n'+k} ~.
\label{PFP}
\end{equation}
The $SU(2)$ $P$ matrix has the following property as
\begin{equation}
 P_j^{~k/2 - j'} = (-1)^{j-j'}  P_{k/2 -j} ^{~ j'} ~,
\end{equation}  
which will be often used in the calculation.

Now we can construct the crosscap states of parafermionic theory.
The crosscap states of (\ref{BS}) are obtained by using parafermionic
$P$ matrix (\ref{PFP}) as
\begin{equation}
 \ket{OA,0,\hat{n}}_C = \sum _{(j,n) \in PF(k)} 
 \frac{P^{PF~j,n}_{0, \hat{n}}}{\sqrt{S_{0,0}^{PF~j,n}}} \ket{OA,j,n}_I ~, 
\label{A-planeI}
\end{equation}
where $\ket{OA,j,n}_I$ are Ishibashi crosscap states.
The $P$ matrices restrict to the label $\hat{n} \in 2 \bz$.
We can also obtain the crosscap states of (\ref{BS2}) by using simple
currents as
\begin{equation}
 \ket{OA,\frac{k}{2},\hat{n}}_C = \sum _{(j,n) \in PF(k)} 
 \frac{P^{PF~j,n}_{\frac{k}{2} ,\hat{n}}}{\sqrt{S_{0,0}^{PF~j,n}}} 
 \ket{OA,j,n}_I ~.
\label{A-planeII}
\end{equation}
However, these crosscap states represent the same ones as
(\ref{A-planeI}) since there is the
spectral flow identification.
Therefore, we only consider the former ones.
The Klein bottle amplitudes can be calculated as
\begin{equation}
 \cK =  ~_C \bra{0A,0,\hat{n}} \tq^{H_c} \ket{0A,0,\hat{n}}_C  
  = \sum_{j =0}^{k/2} \chi_{j,0} (q) +
 E_k (-1)^{\hat{n}/2}\chi_{k/4 , k/2} (q)~.
\end{equation}
The first term has only $n=0$ sector in contrast to the $U(1)$ case
(\ref{U1AKB}) because of the spectral flow identification (\ref{PFI}).
The last term exists since it is the fixed point of the spectral flow.  

The M\"{o}bius strip amplitudes between A-branes and A-orientifolds can
be obtained after some calculation as 
\begin{equation}
 \cM = ~_C \bra{A,\hat{j},\hat{n}} \tq^{H_c} \ket{OA,0,\hat{n}'}_C
 = \sum_{j = 0}^{k/2}   Y_{\hat{j},0}^{~~j}
  \hat{\chi}_{j , 2 \hat{n} - \hat{n}'} (q)~,~
 Y_{\hat{j},0}^{~~j} = 
 (-1)^{2 \hat{j}} (-1)^{j} N^{~~j}_{\hat{j}, \hat{j}} ~.
\label{A-OA}
\end{equation}
The following formula for $SU(2)$ $Y$ matrix is
useful to show the above equation
\begin{equation}
  Y_{i,k/2-j}^{~~k/2-l} = (-1)^{l-j-2i} Y_{i,j}^{~~l} ~. 
\end{equation}
From the above amplitudes (\ref{A-OA}), we can see that the mirror image
of A-brane is $\ket{A,\hat{j},\hat{n}'-\hat{n}}$, in fact, 
\begin{equation}
 \cA = 
~_C \bra{A,\hat{j},\hat{n}} \tq^{H_c} \ket{A,\hat{j},\hat{n}'-\hat{n}}_C
 = \sum_{j = 0}^{k/2}   N_{\hat{j},\hat{j}}^{~~j}
  \chi_{j , 2 \hat{n} - \hat{n}'} (q)~,
\end{equation} 
where the spectrum is the same as in the M\"{o}bius strip amplitudes.
In appendix B, we determine the geometry of A-orientifolds and 
identify them as the combination of $\cO$0-planes at
the opposite boundary points and $\cO$1-planes connecting these points.
The M\"{o}bius strip amplitudes between B-branes and A-orientifolds can
be calculated as 
\begin{equation}
 \cM = ~_C \bra{B,\hat{j}} \tq^{H_c} \ket{OA,0,\hat{n}}_C 
     = \frac{1}{2} \sum_{j=0}^{k/2} 
   \left(Y_{\hat{j},0}^{~~j} + 
 (-1)^{\frac{\hat{n}}{2}} Y_{\hat{j},\frac{k}{2}}^{~~j} \right)
  \hat{\chi} ^{''} _{j} (q) ~,
\label{OA-B}
\end{equation} 
where $ \hat{\chi} ^{''} _{j'} (q)$ are defined by $P$ transformation of 
$\hat{\chi}' _j (q) $ (\ref{chi'}) as
\begin{equation}
 \hat{\chi}' _j (\tq) = \sum_{j'=0}^{k/2}
 P_j ^{~ j'}\hat{\chi}^{''} _{j'} (q) ~,~
 \chi^{''} _{j'} (q) = 
 e^{\pi i k \tau / 2} \chi^{SU(2)} _{j'} (\tau, -\tau) / \chi_{ND} (\tau) ~.
\label{chi''}
\end{equation}

We have already seen that we can construct the other type of boundary
states, which are called as B-type boundary states and are constructed by
using $\bz_k$ orbifold procedure and T-duality.
Just as the crosscap states of $U(1)$ WZW models, we can get the 
B-type crosscap states in parafermionic theory by the same procedure.
The results are given by two type orientifolds, which are called as
B-type and B'-type ones.
First, we concentrate on the B-type ones and they
are given in the even $k$ case as
\begin{equation}
 \ket{OB,0}_C =  
 \left(\frac{k}{2}\right)^{\frac{1}{4}}  \sum_{j=0}^{k/2} 
 \frac{P_0^{~j}}{\sqrt{S^{~j}_{0}}} \ket{OB,j,0}_I ~,
\label{B-planeI}
\end{equation}
where B-type Ishibashi crosscap states are defined in the same way as
B-type Ishibashi boundary states.
The orientifold projection given by this crosscap state can be seen by
the Klein bottle amplitude
\begin{equation}
\cK =  ~_C \bra{OB,0} \tq^{H_c} \ket{OB,0}_C 
 = \sum_{(j,n) \in PF(k)} Y_{0,0}^{~~j} \chi_{j,n} (q) ~.
\end{equation}
The geometry of B-type orientifold can be identified as 
$\cO$0-plane located at the center of the disk, which can be determined by
the scattering with the closed string states as in
appendix B. It can be seen that
this B-type orientifolds originate from the $\cO$0-plane parts of
A-orientifolds. Thus the geometrical description of 
$\bz_k$ orbifold and T-duality gives the same result. 
The M\"{o}bius strip amplitudes with the boundary states 
can be obtained just as A-type crosscap states.
The amplitudes with B-type boundary states are given by
\begin{equation}
 \cM = ~_C \bra{B,\hat{j}} \tq^{H_c} \ket{OB,0}_C = 
 \sum_{(j,n) \in PF(k)}  Y_{\hat{j},0}^{~~j} 
   \hat{\chi}_{j,n} (q) ~,
\end{equation}
and the amplitudes with A-type boundary states are obtained as
\begin{equation}
  \cM = ~_C \bra{A,\hat{j}, \hat{n}} \tq^{H_c} \ket{OB,0}_C = 
 \sum_{j=0}^{k/2} Y_{\hat{j},0}^{~~j} 
   \hat{\chi}^{''}_{j} (q) ~,
\end{equation}
where we use the character $\hat{\chi}^{''}_{j} (q)$ defined above 
(\ref{chi''}).

We can also obtain the other (B'-type) crosscap state by $\bz_k$
orbifold procedure and T-duality and this state is given by
\begin{equation}
 \ket{OB,\frac{k}{2}}_C = 
 \left(\frac{k}{2}\right)^{\frac{1}{4}}  \sum_{j=0}^{k/2} 
 \frac{P_{\frac{k}{2}}^{~j}}{\sqrt{S^{~j}_{0}}} \ket{OB,j,0}_I ~.
\label{B-planeII}
\end{equation}
The Klein bottle amplitude between this crosscap state is
\begin{equation}
\cK =  ~_C \bra{OB,\frac{k}{2}} \tq^{H_c} \ket{OB,\frac{k}{2}}_C 
 = \sum_{(j,n) \in PF(k)} 
  Y_{\frac{k}{2},\frac{k}{2}}^{~~j} \chi_{j,n} (q) ~.
\end{equation}
The geometry of this orientifold is also determined by scattering with
the closed string states in appendix B. There we show that this
orientifold is $\cO$2-plane wrapping the whole disk.
The B'-type orientifolds originate from the $\cO$1-plane parts of
the A-orientifolds. Thus, summing all images and performing 
T-duality gives the same geometrical interpretation.
The M\"{o}bius strip amplitudes with B-type branes are given by
\begin{equation}
 \cM = ~_C \bra{B,\hat{j}} \tq^{H_c} \ket{OB,\frac{k}{2}}_C = 
 \sum_{(j,n) \in PF(k)}
  Y_{\hat{j},\frac{k}{2}}^{~~j} 
   \hat{\chi}_{j,n} (q) ~,
\end{equation}
and the amplitudes with A-type branes are obtained as
\begin{equation}
  \cM = ~_C \bra{A,\hat{j}, \hat{n}} \tq^{H_c} \ket{OB,\frac{k}{2}}_C = 
 \sum_{j=0}^{k/2} Y_{\hat{j},\frac{k}{2}}^{~~j} 
   \hat{\chi}^{''}_{j} (q) ~,
\end{equation}
where we again use the character (\ref{chi''}).

%%%%%%%%%%%%%%%%%%%%%%%%%%%%%%%%%%%%%%%%%%%%%%%%%%%%%%%%%%%%%%%%%%%%%

\section{Super $SU(2)/U(1)$ WZW Models and Orientifolds}
\indent

In order to apply to superstring theories,
we have to extend the previous analysis to the supersymmetric one.
Although $\cN = 1$ supersymmetry is included in super parafermionic
theory, the supersymmetry is enhanced to $\cN=2$.
We study the closed strings and D-branes in super
parafermionic theory in the next subsection and the orientifolds in subsection
5.2. The construction of orientifolds can be proceeded just as the bosonic 
case and we get the similar results.

\subsection{Super $SU(2)/U(1)$ WZW models and D-branes}
\indent

This theory is known as $\cN=2$ minimal model and is given by  
$(SU(2)_k \times U(1)_2 ) / U(1)_{k+2}$ coset WZW models.
The Hilbert space of super parafermionic theory is given by the
decomposition as
\begin{equation}
 \cH_{j,n}^{SU(2)} \otimes \cH_s^{U(1)_2} = 
 \cH_{j,n,s}^{SPF} \otimes \cH_{n}^{U(1)_{k+2}} ~,
\end{equation}
where the labels run $j = 0, 1/2, \cdots, k/2$, $n \in \bz_{2k+4}$ and 
$s \in \bz_4$. 
The states with $s = 0, 2$ are in the NS sector and the ones with $s = 1,3$
are in the R  sector. 
Just as the bosonic case, there are the spectral flow
identification
\begin{equation}
 \cH_{j,n,s}^{SPF} = \cH_{\frac{k}{2}-j, n+k+2, s+2}^{SPF} ~,
\label{SPFI}
\end{equation}
and the restriction of the label $2j + n + s \in 2 \bz$.
We will call the distinct irreducible representation of
super parafermionic theory as $SPF(k)$ and the characters as 
$\chi_{j,n,s} (q)$.
The lowest conformal weights are given by
\begin{equation}
 h_{j,n,s} = \left\{
\begin{array}{ll}
 \frac{j(j+1)}{k+2} - \frac{n^2}{4(k+2)} + \frac{s^2}{8}
 & -2j \leq n -s  \leq 2j \\
 \frac{j(j+1)}{k+2} - \frac{n^2}{4(k+2)} + \frac{s^2}{8} 
 + \frac{n-s-2j}{2} &
 2j \leq n-s \leq 2k-2j ~,
\end{array}
\right. 
\label{SPFcw}
\end{equation}
and the conformal weights of the fields outside this region can be
obtained by making use of the spectral flow.
The modular transformation matrix is given by
\begin{equation}
 S_{j,n,s}^{SPF~j',n',s'} = \frac{1}{k+2} \sin \left(
 \frac{\pi (2j+1)(2j'+1)}{k+2}
\right) e^{\pi i n n'/(k+2)} e^{-\pi i s s' /2} ~,
\end{equation}
and the fusion coefficients are 
\begin{equation}
 N^{SPF~(j^{''},n^{''},s^{''})}_{(j,n,s),(j',n',s')}
 = N_{j,j'}^{~j^{''}} \delta^{(2k+4)}_{n+n'-n^{''}} 
 \delta^{(4)}_{s+s'-s^{''}} + 
   N_{j,j'}^{~~k/2 -j^{''}} \delta^{(2k+4)}_{n+n'-n^{''}-k-2} 
 \delta^{(4)}_{s+s'-s^{''}-2} ~.
\end{equation}
The diagonal modular invariant is given by
\begin{equation}
 \cT = \sum_{(j,n,s) \in SPF(k)} |\chi_{j,n,s} (q)|^2 ~,
\label{SPFDMI}
\end{equation}
and this theory has the global symmetries which are generated by
\begin{eqnarray}
  g_1 \Phi_{j,n,s} &=& e^{2 \pi i (n /(2k+4) - s/4)} \Phi_{j,n,s} \nn
  g_2 \Phi_{j,n,s} &=& e^{\pi i s} \Phi_{j,n,s} ~.
\end{eqnarray}
We will use the orbifold procedure as $\bz_{k+2} \times \bz_2$
symmetry, which is generated by $g_1^2 g_2 \times g_2$.

Let us study D-branes in super parafermionic theory.
The A-type Cardy states can be constructed in the usual way as
\begin{equation}
 \ket{A,\hat{j},\hat{n},\hat{s}} _C
 = \sum_{(j,n,s) \in SPF(k)}
  \frac{S^{SPF~j,n,s}_{\hat{j},\hat{n},\hat{s}}}
 {\sqrt{S^{SPF~j,n,s}_{0,0,0}}} \ket{A,j,n,s}_I ~,
\end{equation}
where $\ket{A,j,n,s}_I$ are Ishibashi states in super parafermionic
theory. 
The geometric picture of A-branes is given as follows.
The target space of super parafermionic theory is the disk with fermions and
it is convenient to use the disk with $2k+4$ special points by making
use of the bosonized fermions \cite{MMS}. Then the geometry of D-brane
described by $\ket{A,\hat{j},\hat{n},\hat{s}}$ is determined as D1-brane, 
which is the straight line connecting $4\hat{j} + 2$ separated points.  
The label $\hat{n}$ can be changed by the $\bz_{k+2}$ rotation of the disk.
The label $\hat{s}$ means the spin structure and the states with 
$\hat{s}$ and $\hat{s}+2$ describes the D1-branes with opposite
orientation.  
The annulus amplitudes are given by  
\begin{eqnarray}
\cA = {}_C\bra{A,\hat{j},\hat{n},\hat{s}} \tq^{H_c} 
 \ket{A,\hat{j}',\hat{n}',\hat{s}'}_C
 &=& \sum_{(j,n,s) \in SPF(k)} 
 N^{SPF~~~~~~(j,n,s)}_
  {(\hat{j},-\hat{n},-\hat{s}),(\hat{j}',\hat{n}',\hat{s}')} 
 \chi_{j,n,s} (q) \nn
 &=& \sum_{j=0}^{k/2} N^{~~j} _{\hat{j},\hat{j}'} 
 \chi_{j,\hat{n}-\hat{n}',\hat{s}-\hat{s}'} (q) ~.
\end{eqnarray}
We should note that the bra state ${_C}\bra{A,\hat{j},\hat{n},\hat{s}+2}$ has  
the same orientation with $ \ket{A,\hat{j},\hat{n},\hat{s}} _C$.
We can see that there is no tachyon in the open string
spectrum of the annulus amplitudes between the same branes.

The B-type Cardy states are similarly constructed by using $\bz_{k+2} \times
\bz_2$ orbifold and T-duality. 
The orbifold procedure restricts the labels of Ishibashi states to
 $n=0,k+2$ and $s=0,2$ and we will only use $n=0$ states by using the
spectral flow identification (\ref{SPFI}).
Performing T-duality converts from A-type Ishibashi states to B-type ones.
The A-type Ishibashi states can be obtained by the decomposition 
\begin{equation}
 \ket{j}^{SU(2)_k}_I \otimes \ket{A,s}_I^{U(1)_2}
 = \sum_{n=0}^{2k+3} \ket{A,j,n,s}_I^{SPF} \otimes \ket{A,n}^{U(1)_{k+2}}_I ~,
\label{decompose}
\end{equation}
and we can apply T-duality just as the bosonic case by making use of the
operator $\exp (\pi i J^1_0)$  as
\begin{eqnarray}
 \lefteqn{(1 \otimes e^{\pi i \tilde{J}^1_0})
 \ket{j}^{SU(2)_k}_I \otimes \ket{B,s,-s}_I^{U(1)_2}} \nn
 && \hspace{2cm} =\sum_{n=0}^{2k+3} \ket{B,j,n,-n,s,-s}_I^{SPF} 
 \otimes \ket{B,n,-n}^{U(1)_{k+2}}_I ~,
\end{eqnarray}
where $\ket{B,r,-r}^{U(1)}_I$ are defined by (\ref{Br-r}).

Now we can construct the B-type Cardy boundary  states as
\begin{equation}
 \ket{B,\hat{j},\hat{n},\hat{s}}_C 
 = \sqrt{2(k+2)} \sum_{j \in \bz} \sum_{s=0,2}
 \frac{S^{SPF~ j,0,s}_{\hat{j},\hat{n},\hat{s}}}
      {\sqrt{S^{SPF~j,0,s}_{0,0,0}}} \ket{B,j,0,0,s,-s}_I ~,
\end{equation}
and these states represent D2-branes at the center of the disk.
Since we sum over only $s=0,2$, the label of the distinct states is
given by $\hat{s}=0,1$, which corresponds to the spin structure. 
The label $\hat{n}$ is only used to be survived
by the selection rule $2\hat{j}+\hat{n}+\hat{s} \in 2 \bz$.
Therefore there are $\hat{j} = 0, 1/2, \cdots, (k-1)/4$ states with odd $k$.
In even $k$ case, there is special $\hat{j}=k/4$ state in addition to 
 $\hat{j} = 0, 1/2, \cdots, (k-2)/4$ states. 
The general $\hat{j}$ states are non-orientable
and do not have RR charges. 
In fact, they are unstable because there
are tachyons in the open string spectrum of the
annulus amplitudes between B-type boundary states 
\begin{equation}
\cA = {}_C \bra{B,\hat{j},\hat{n},\hat{s}} \tq^{H_c} 
 \ket{B,\hat{j}',\hat{n}',\hat{s}'}_C 
 = \sum_{(j,n,s) \in SPF(k)} E_{\hat{s}' - \hat{s} -s} 
 (N_{\hat{j},\hat{j}'}^{~~j} + N_{\hat{j},\hat{j}'}^{~~k/2-j})
  \chi_{j,n,s} (q)
~.
\end{equation} 
On the contrary, the state with
$\hat{j}=k/4 $ is orientable and has RR charge, moreover it is free
of tachyons in the open string amplitudes between this type of boundary
states. 
The detailed analysis was given in \cite{MMS}, and we do not deal
with this state for simplicity.  

The annulus amplitudes between A-type and B-type boundary states 
can be obtained by using the overlaps between A-type and B-type 
Ishibashi states as
\begin{equation}
  {}_I \bra{A,j,0,0} \tq^{H_c} \ket{B,j',0,0}_I 
 = \delta_{j,j'}
  \chi^{SU(2)}_j (-1/\tau, 1/2) ~. 
\end{equation} 
The character $\chi_{ND}$ as in (\ref{chi'}) does not appear since there
are the $U(1)$ sectors in the both sides of the equation (\ref{decompose}).
Thus the amplitudes are given by
\begin{equation}
\cA = {}_C \bra{A,\hat{j},\hat{n},\hat{s}} \tq^{H_c} 
 \ket{B,\hat{j}',\hat{n}',\hat{s}'} _C
 = \sum_{j=0}^{k/2} N_{\hat{j},\hat{j}'}^{~~j} 
  e^{\pi i k \tau / 8} \chi^{SU(2)}_j (\tau,\tau/2)
~,
\end{equation} 
which are independent with the labels $\hat{n},\hat{n}',\hat{s}$ and
$\hat{s}'$.

\subsection{Crosscap states in Super $SU(2)/U(1)$ WZW Models}
\indent

The crosscap states in super parafermionic theory can be constructed
just like the bosonic case by using the $P$ matrix 
\begin{equation}
 P^{SPF~j',n',s'}_{j,n,s} = 
 P_j^{~j'} P_n^{~n'} P_s^{~s'} +
 (-1)^{(2j-n+s)/2}   
 P_j^{~k/2-j'} P_n^{~n'+k+2} P_s^{~s'+2} ~,
\label{SPFP}
\end{equation}  
where we used the $U(1)$ $P$ matrices as
\begin{equation}
  P_n^{~n'} = \frac{1}{\sqrt{k+2}} e^{\pi i n n' /2(k+2)} 
     E_{n + n' + k + 2}~,~~
  P_s^{~s'}  = \frac{1}{\sqrt{2}} e^{- \pi i s s' /4} E_{s + s' + 2} ~.
\end{equation}
The last term of (\ref{SPFP})
is included because of the spectral flow identification just as
(\ref{PFP}). 
The A-type orientifolds can be obtained by the solution (\ref{BS}) as
\begin{equation}
 \ket{OA,0,\hat{n},\hat{s}}_C 
  = \sum_{(j,n,s) \in SPF(k)} 
 \frac{P^{SPF~j,n,s}_{0,\hat{n},\hat{s}}}
 {\sqrt{P^{SPF~j,n,s}_{0,0,0}}} \ket{OA,j,n,s}_I ~, 
\end{equation}
where $\ket{OA,j,n,s}_I$ are Ishibashi crosscap states in super
parafermionic theory.
These states represent $\cO$1-planes connecting the opposite points of the
boundary. 
By applying the simple current technique, we can obtain the solution
(\ref{BS2}) 
\begin{equation}
 \ket{OA,\frac{k}{2},\hat{n},\hat{s}}_C 
 = \sum_{(j,n,s) \in SPF(k)} 
 \frac{P^{SPF~j,n,s}_{\frac{k}{2},\hat{n},\hat{s}}}
 {\sqrt{P^{SPF~j,n,s}_{0,0,0}}} \ket{OA,j,n,s}_I ~, 
\end{equation}
which are the same as the previous ones just as the bosonic case.
We can calculate the Klein bottle amplitudes as
\begin{eqnarray}
\cK &=&  {}_C \bra{OA,0,\hat{n},\hat{s}+4} \tq^{H_c} 
 \ket{OA,0,\hat{n},\hat{s}}_C \nn
  &=& \sum_{j=0}^{k/2} 
 (\chi_{j,0,2} (q) + (-1)^{\hat{s}} \chi_{j,0,0} (q)) \nn
 & &  
  +~ E_k (-1)^{(\hat{n} - \hat{s})/2}
 ( \chi_{\frac{k}{4}, \frac{k+2}{2}, 3} (q) 
 + (-1)^{\hat{s}} \chi_{\frac{k}{4}, \frac{k+2}{2}, 1} (q) ) ~.
\label{SOA-OA}
\end{eqnarray}
We should note here that the bra state
$\bra{OA,\hat{j},\hat{n},\hat{s}+4}_C$ corresponds to the same orientifold
described by the ket state $\ket{A,\hat{j},\hat{n},\hat{s}}_C$.
If we apply to superstring theories, one might be able to remove tachyons by
performing appropriate GSO projection.
The M\"{o}bius strip amplitudes with A-type branes are given by 
\begin{equation}
\cM =  {}_C \bra{A,\hat{j},\hat{n},\hat{s}} \tq^{H_c}
 \ket{OA,0,\hat{n}',\hat{s}'}_C
 = \sum_{j=0}^{k/2} Y_{\hat{j},0}^{~~j} 
 \hat{\chi}_{j , 2 \hat{n} - \hat{n}' , 2 \hat{s}-\hat{s}'} (q) ~,
\end{equation}
and the amplitudes with B-branes are obtained as
\begin{equation}
\cM =  {}_C \bra{B,\hat{j},\hat{n},\hat{s}} \tq^{H_c}
 \ket{OA,0,\hat{n}',\hat{s}'}_C
 = \frac{1}{2} \sum_{j=0}^{k/2}
 \left(  Y_{\hat{j},0}^{~~j} + (-1)^{\frac{\hat{n}' -\hat{s}'}{2}}
Y_{\hat{j},\frac{k}{2}}^{~~j} 
 \right)
  \hat{\chi}_j ^{'''} (q) ~,
\end{equation}
where the characters are defined just like the bosonic case
(\ref{chi''}) as
\begin{equation}
 \chi^{'''}_j (q)  = e^{\pi i k \tau /2 } \chi^{SU(2)}_j (\tau,-\tau) ~.
\label{chi'''}
\end{equation}

The construction of B-type orientifolds can be made by $\bz_{k+2}
\times \bz_2$ orbifold and T-duality as before.
Performing T-duality makes B-type Ishibashi crosscap states from A-type
ones just as the B-type boundary states. 
B-Type Cardy crosscap state is given by (only for even $k$)
\begin{equation}
 \ket{OB,0}_C = (2k+4)^{1/4} \sum_{j=0}^{k/2} 
 \frac{P_0^{~j}}{\sqrt{S_{0}^{~j}}} \ket{OB,j,0,0}_I ~,
\end{equation} 
which is the $\cO$0-plane at the center of the disk.
The orientifold projection is given by taking the right-mover of the
diagonal modular invariant (\ref{SPFDMI}) with phases as
\begin{equation}
\cK =  {}_C \bra{OB,0} \tq^{H_c} \ket{OB,0}_C = 
 \sum_{(j,n,s) \in SPF(k)} Y_{0,0}^{~~j} \chi_{j,n,s} (q) ~.
\end{equation}
There are tachyons in the direct channel spectrum of the amplitude.
These tachyons are related to the fact that we are dealing with the
system with the diagonal modular invariant (\ref{SPFDMI}).
Therefore if we apply the correct GSO projection, we can remove these
tachyons and make the system stable.
Furthermore we can calculate the  M\"{o}bius strip amplitudes with the
boundary states in super parafermionic theory. 
The M\"{o}bius strip amplitudes with B-type branes can be given by
\begin{equation}
\cM =  {}_C \bra{B,\hat{j},\hat{n},\hat{s}} \tq^{H_c} \ket{OB,0}_C = 
 \sum_{(j,n,s) \in SPF(k)} Y_{\hat{j},0}^{~~j} \hat{\chi}_{j,n,s} (q) ~,
\end{equation}
and the amplitudes with A-type branes are obtained by using the
characters $(\ref{chi'''})$ as
\begin{equation}
\cM  = {}_C \bra{A,\hat{j},\hat{n},\hat{s}} \tq^{H_c} 
 \ket{OB,0}_C  = \sum_{j=0}^{k/2} Y_{\hat{j},0}^{~~j}
 \hat{\chi}_j ^{'''} (q) ~. 
\end{equation} 

The B'-type orientifolds can be also constructed by $\bz_{k+2} \times \bz_2$
orbifold and T-duality and we obtain
\begin{equation}
 \ket{OB,\frac{k}{2}}_C = (2k+4)^{1/4} \sum_{j=0}^{k/2} 
 \frac{P_{\frac{k}{2}}^{~j}}{\sqrt{S_0^{~j}}} \ket{OB,j,0,0}_I ~,
\end{equation} 
which is the $\cO$2-plane wrapping  the disk.
The orientifold projection is given by
\begin{equation}
\cK =  {}_C \bra{OB,\frac{k}{2}} \tq^{H_c} \ket{OB,\frac{k}{2}}_C = 
 \sum_{(j,n,s) \in SPF(k)} Y_{\frac{k}{2},\frac{k}{2}}^{~~j} 
\chi_{j,n,s} (q) ~,
\end{equation}
which is the right-mover of the diagonal modular invariant
(\ref{SPFDMI}) and 
this may be the most natural orbifold projection.
The M\"{o}bius strip amplitudes with B-type boundary states are given by
\begin{equation}
\cM =  {}_C \bra{B,\hat{j},\hat{n},\hat{s}} \tq^{H_c} \ket{OB,\frac{k}{2}}_C = 
 \sum_{(j,n,s) \in SPF(k)} Y_{\hat{j},\frac{k}{2}}^{~~j} 
\hat{\chi}_{j,n,s} (q) ~,
\end{equation}
and the amplitudes with A-type branes are obtained as
\begin{equation}
\cM = {}_C \bra{A,\hat{j},\hat{n},\hat{s}} \tq^{H_c} 
 \ket{OB,\frac{k}{2}}_C  = \sum_{j=0}^{k/2} Y_{\hat{j},\frac{k}{2}}^{~~j}
  \hat{\chi}_j^{'''} (q) ~. 
\end{equation}

%%%%%%%%%%%%%%%%%%%%%%%%%%%%%%%%%%%%%%%%%%%%%%%%%%%%%%%%%%%%%%%%%%%%%
%\newpage
\section{Conclusion}
\indent

In this paper, the orientifolds of parafermionic theories are
investigated. The D-branes in parafermionic theories were analyzed in
\cite{MMS} and they determined the geometry of D-branes by scattering
with the closed string states.
The geometry of orientifolds can be also determined by scattering the
closed string states and we summarize in appendix B.
The paper \cite{MMS} also constructed the new type of branes
called as B-type branes. In $SU(2)_k/U(1)_k$ WZW model, it can be seen
that the $\bz_k$ orbifold is T-dual to the original theory.
Therefore we can construct new type branes from the previously known
branes called as Cardy type branes or A-type branes.  
The same method can be applied to the orientifolds of parafermionic
theory.  
The Cardy type orientifolds, called as A-type orientifolds, were
constructed in \cite{sagnotti1,sagnotti2,sagnotti3,sagnotti4}.
In our case, there are A-type crosscap states (\ref{A-planeI})
which are the solutions of (\ref{BS}). They are identified as the
$\cO$1-planes connecting the boundary points.
Using $\bz_k$ orbifold and T-duality, we can construct B-type
and B'-type orientifolds.
B-type crosscap state is given by (\ref{B-planeI}) and it represents
$\cO$0-plane at the center of the disk.
B'-type crosscap state is given by (\ref{B-planeII}) and it represents 
$\cO$2-plane wrapping the whole disk.
The extension to the supersymmetric case is also discussed in section 5. 

It is known that the D2-branes in $SU(2)$ WZW models can be
thought of the bound states of D0-branes
\cite{alek1,bds,paw,alek2,hk,hikida}.
This fact can be used to define the D-brane charges and they can be
classified by twisted K-theory \cite{BM,FS,MMS2,isidro,evslin}. 
Therefore it is interesting to calculate the K-theoretic D-brane charges
in the system with orientifolds of WZW models just like flat space analysis 
\cite{sugimoto}.     
The extension to the orientifolds of other backgrounds are also 
interesting, for example, more general coset WZW models.
The study of orientifolds of $AdS_3$ space seems important because the
orientifolds wrap the twisted conjugacy classes as pointed out in
\cite{oplane3}. 
We might be able to find the orientifolds wrapping the twisted
conjugacy classes in $SU(N)$ WZW models.

%%%%%%%%%%%%%%%%%%%%%%%%%%%%%%%%%%%%%%%%%%%%%%%%%%%%%%%%%%%%%%%%%%%%%

\section*{Acknowledgement}
\indent

We would like to thank T. Eguchi,  M. Nozaki, Y. Sugawara, T. Takayanagi
and M. Yamada for useful discussions and comments.

%%%%%%%%%%%%%%%%%%%%%%%%%%%%%%%%%%%%%%%%%%%%%%%%%%%%%%%%%%%%%%%%%%%%%%

\section*{Appendix A ~ Modular Transformations of Several Functions}
\setcounter{equation}{0}
\def\theequation{A.\arabic{equation}}
\indent

The Dedekind $\eta$ function is given by
\begin{equation}
 \eta (\tau) = q^{\frac{1}{24}} \prod_{n=1}^{\infty} (1 - q^n) ~,
\end{equation}
where $q=\exp (2 \pi i \tau)$ and its modular transformation is
\begin{equation}
 \eta (\tau + 1) =
  e^{\pi i / 12} \eta (\tau) ~,~~
 \eta (- \frac{1}{\tau}) =
  \sqrt{-i \tau} \eta (\tau) ~.
\label{etam}
\end{equation}

The theta functions are defined as
\begin{equation}
 \Theta_{n,k} (\tau,z) = \sum_{l \in \bz} 
     e^{2 \pi i \tau k \left(l + \frac{n}{2k}\right)^2}
    e^{2 \pi i z k \left( l + \frac{n}{2k}\right)} ~,
\end{equation}
and their modular transformations are
\begin{eqnarray}
 \Theta_{n,k} (\tau + 1 , z) &=&
  e^{\pi i n^2/2k } \Theta_{n,k} (\tau,z) \nn
 \Theta_{n,k} (- \frac{1}{\tau} , - \frac{z}{\tau}) &=&
  \sqrt{\frac{-i \tau}{2k}} e^{2 \pi i k z^2 /4 \tau}
  \sum_{n'} e^{- i \pi n n' /k} \Theta_{n' , k} (\tau ,z) ~.
\end{eqnarray} 

The characters of $SU(2)_k$ current algebra are written by using the theta
functions as
\begin{equation}
 \chi^{SU(2)}_j (\tau , z) =
  \frac{\Theta_{2j+1,k+2} - \Theta_{-(2j+1), k+2}}
                   {\Theta_{1,2} - \Theta_{-1,2}} (\tau ,z) ~,
\end{equation}
 and their modular transformations are
\begin{eqnarray}
 \chi^{SU(2)}_j (\tau + 1 , z) &=&
  e^{ 2 \pi i \left( \frac{j(j+1)}{k+2} - \frac{k}{8(k+2)}\right) }
     \chi^{SU(2)}_j (\tau,z) \nn
 \chi^{SU(2)}_j (- \frac{1}{\tau} , - \frac{z}{\tau}) &=&
   e^{2 \pi i k z^2 /4 \tau}
  \sum_{j'=0}^{k/2} S^{~j'}_{j} \chi^{SU(2)}_{j'} (\tau ,z) ~,
\end{eqnarray} 
where $S$ matrix of $SU(2)$ character are given as
\begin{equation}
 S^{~j'}_{j} = \sqrt{\frac{2}{k+2}} 
 \sin \left( \frac{\pi (2 j + 1)(2 j' + 1) }{k+2}\right) ~.
\end{equation} 

%%%%%%%%%%%%%%%%%%%%%%%%%%%%%%%%%%%%%%%%%%%%%%%%%%%%%%%%%%%%%%%%%%%%%%%%%

\section*{Appendix B ~ Geometry of Orientifolds}
\setcounter{equation}{0}
\def\theequation{B.\arabic{equation}}
\indent

Let us first examine the geometry of $SU(2)_k$ WZW model, whose target
space is $S^3$.
The metric of $S^3$ can be parametrized in the term of Euler angle
\begin{equation}
 g = e^{ i \chi \frac{\sigma^3}{2}} 
      e^{ i \theta \frac{\sigma^1}{2}}  e^{ i \varphi \frac{\sigma^3}{2}} ~,
\label{euler}
\end{equation}
 where $0 \leq \theta \leq \pi$, $0 \leq \varphi \leq 2 \pi$ and $0 \leq
 \chi \leq 4 \pi$.
When analyzing D-branes in $SU(2)_k$ WZW models, it is convenient to use
 the metric
\begin{equation}
 ds^2 = d \psi^2 + \sin^2 \psi d^2 s_{S^2} ~,
\label{psi}
\end{equation}
because the maximally symmetric D-branes are located at $\psi =$ const. 

The parafermionic theory can be obtained by gauging $U(1)$ sector of
$SU(2)$ currents. In the term of Euler angle (\ref{euler}), this $U(1)$
sector corresponds to the shift of $\chi + \varphi$. Using
$\rho = \sin \frac{\theta}{2}$ and $\phi = \frac{1}{2}(\chi - \varphi)$, 
the target space of parafermionic theory
can be described as the disk with the radius $R=1$;
\begin{eqnarray}
 d s^2 &=& \frac{k}{1 - \rho^2} (d \rho ^2 + \rho ^2 d \phi^2) \nn
 g_s &=& g_s (0) (1- \rho^2)^{-\frac{1}{2}} ~.
\end{eqnarray}
T-duality can be performed on the $U(1)$ isometry and we get
 \begin{eqnarray}
 d s^2 &=&
  \frac{k}{1 - {\rho '}^2} (d {\rho '} ^2 + {\rho '}^2 d {\phi '}^2) \nn
 g_s &=& \frac{g_s (0)}{\sqrt{k}} (1- {\rho'}^2)^{-\frac{1}{2}} ~,
\end{eqnarray}
where ${\rho '}^2 = (1 - \rho^2)$ and $ \phi ' \sim \phi ' + 2 \pi / k$.
Thus we can see that the original theory is T-dual to $\bz_k$ orbifold and
the center of the disk of the original theory corresponds to
the boundary of the disk.

The geometry of D-branes and orientifolds can be determined by scattering
with the closed strings \cite{FFFS,MMS,oplane1} in the large $k$ limit.
Since the shapes of D-branes in parafermionic theory were investigated in
\cite{MMS}, we concentrate on the shapes of orientifolds.  
By using the matrix representation
\begin{equation}
 \cD^j_{mm'} (g) = \bra{j,m} e^{ i \chi J^3_0} 
      e^{ i \theta J^1_0}  e^{ i \varphi J^3_0} \ket{j,m'} ~,~
  \bra{ j,m} j,m' \rangle = \delta_{m,m'} ~,
\end{equation}
we define the closed string states $\ket{g}$ which have the
properties as
\begin{equation}
 \cD^j _{m,m'} (g) \sim \frac{1}{\sqrt{2j+1}} \bra{g} j,m,m' \rangle ~.
\end{equation}
These closed string states are well localized if we use $j \leq k/2$.
As we said above, the geometry of parafermionic theory can be obtained
by gauging $\chi + \varphi$. Thus we scatter the closed string states
$\ket{\theta,\phi}$, which satisfy
\begin{equation}
 e^{i \phi m} \bra{j,m} e^{i \theta J^1_0} \ket{j,-m} = 
 \frac{1}{\sqrt{2j+1}} \bra{\theta,\phi} j,m,m \rangle ~.     
\end{equation}

Now we can examine the geometry of orientifolds.
We begin with A-type orientifolds  (\ref{A-planeI}) in
parafermionic theory. 
We use the states with $\hat{n}=0$ for simplicity and the other
states can be obtained by $\bz_k$ rotation of the disk. 
The scattering amplitude between the A-type crosscap state and the 
closed string state is made of two contributions.
One contribution originates from the first term of (\ref{PFP}) and it is
given by 
\begin{eqnarray}
\lefteqn{ \sum_{j} \sum_{m=-2j}^{2j} P_{0}^{~j}  
 \langle j,m,m \ket{\theta,\phi} }\nn
 &\sim&  \sum_{j} \sum_{m=-2j}^{2j} \sin {[}\hat{\psi}(2j+1) {]} 
   e^{i \phi m} \bra{j,m} e^{i \theta J^1_0} \ket{j,-m} \nn
 &=&  \sum_{j} \sin {[}\hat{\psi}(2j+1) {]} 
    \frac{\sin(2j+1)\psi}{\sin \psi} ~,
\end{eqnarray} 
where we define
\begin{equation}
 \hat{\psi} = \frac{\pi}{2(k+1)} ~,~ ~
 \cos \psi (\theta,\phi) = \cos \phi \sin \frac{\theta}{2} ~.
\end{equation}
Evaluating the summation for small $j$, we get
\begin{equation}
  ~_C \bra{OA,0} \theta,\phi \rangle \sim
 \delta ( \hat{\psi} - \psi (\theta,\phi) ) ~,
\label{G-A-plane}
\end{equation}
therefore we can see that this part corresponds to the 
$\cO$0-planes at $\phi = 0$ and $\phi = \pi$ in the large $k$ limit.
The contribution of the second term (\ref{PFP}) can be obtained by   
replacing $P_0^{~j}$ with $P_{\frac{k}{2}}^{~j}$ and rotating the disk by
the amounts of $\hat{n}=k$. Then, we get the similar result (\ref{G-A-plane})
except for $\hat{\psi} = \frac{\pi (k+1)}{2(k+2)} \to \frac{\pi}{2}$. 
Thus this part corresponds to the $\cO$1-plane on
the line segment between $\phi = 0$ and $\phi = \pi$.
By combining two parts, we can see that the crosscap state
(\ref{A-planeI}) describes $\cO$1-plane connecting the opposite boundary
points of the disk. 
The D-branes in the compact WZW models with finite $k$ are not exactly
localized and are smeared out \cite{FFFS}.
However the orientifolds are obtained by the combination of $\bz_2$
orbifold and worldsheet parity transformation, hence the
orientifolds should be sharply localized even in finite $k$.
This is the remarkable contrast to the D-branes in finite $k$.

The geometry of B-type crosscap state (\ref{B-planeI}) can be also
determined by scattering with closed strings.
Using the Legendre polynomials $ P_j (\cos \theta)$ as
\begin{equation}
\cD_{00}^{~~j} = \bra{j,0}e^{i \theta J^1_0 }\ket{j,0} 
= P_j (\cos \theta) ~,
\end{equation}
we can get 
\begin{eqnarray}
 \bra{OB,0} \theta, \phi \rangle 
 &\sim& \sum_j  \cD_{00}^{~~j} P_0^{~j}  
 \sim  - i e ^{i \hat{\psi}} \sum_j 
  P_j (\cos \theta) e^{i n 2 \hat{\psi}} + c.c. \nn
 &\sim& \frac{\Theta(\cos \theta - \cos 2 \hat{\psi})}
 {\sqrt{\cos \theta - \cos 2 \hat{\phi}}} ~,
\label{G-B-planeI}
\end{eqnarray}
where $\hat{\psi}=\frac{\pi}{2(k+1)}$ and $c.c.$ means complex conjugate.
We also used $\Theta (z)$ as the usual theta function (1 for $z \geq0$ and 0
for others). 
The last line in the above equation can be obtained by using the
generating function of Legendre polynomials
\begin{equation}
 \sum_{n=0} t^n P_n (x) = \frac{1}{\sqrt{1 - 2 t x + t^2}} ~.
\end{equation} 
In the large $k$ limit, $\hat{\psi}$ goes to zero, thus we can conclude
that the B-type orientifold (\ref{B-planeI}) is $\cO$0-plane located
at the center of the disk $(\rho = \sin \frac{\theta}{2} = 0)$.
The shape of B'-orientifold can be determined in the similar way.
The only thing we have to do is replacing $P_{0}^{~j}$ with
$P_{\frac{k}{2}}^{~j}$, hence we get (\ref{G-B-planeI}) with
$\hat{\psi} = \frac{\pi (k+1)}{2 (k+1)} \to \frac{\pi}{2}$.
Thus the B'-type orientifold (\ref{B-planeII}) can be identified as
$\cO$2-plane wrapping the whole disk.

%%%%%%%%%%%%%%%%%%%%%%%%%%%%%%%%%%%%%%%%%%%%%%%%%%%%%%%%%%%%%%%%%%%%%%%%%
%\newpage

%\small

\end{document}